\documentclass[usegraphicx,usenatbib]{mn2e}
\usepackage{psfig}
\renewcommand\[{\begin{equation}}
\renewcommand\]{\end{equation}}

\def\Reff{R_{\rm e}}

\def\nh{{\cal V}}

\catcode`\@=11
\def\gsim{\ifmmode{\mathrel{\mathpalette\@versim>}}
    \else{$\mathrel{\mathpalette\@versim>}$}\fi}
\def\lsim{\ifmmode{\mathrel{\mathpalette\@versim<}}
    \else{$\mathrel{\mathpalette\@versim<}$}\fi}
\def\@versim#1#2{\lower 2.9truept \vbox{\baselineskip 0pt \lineskip
    0.5truept \ialign{$\m@th#1\hfil##\hfil$\crcr#2\crcr\sim\crcr}}}
\catcode`\@=12

\begin{document}

   \title{On the use of X-rays to determine dynamical properties
    of Elliptical galaxies}

   \author[Ciotti \& Pellegrini]
          {Luca Ciotti \& Silvia Pellegrini
           \\
		   Astronomy Department, University of Bologna, 
                       via Ranzani 1, 40127 Bologna, Italy
          }

   \date{Accepted by MNRAS, January 27, 2004}

   \maketitle

\begin{abstract} 
The extended and X-ray emitting interstellar medium of early type
galaxies is often used as a tool to determine their total mass $M$ 
and stellar orbital anisotropy $\beta$ profiles,
based on the hypothesis of hydrostatic equilibrium for the hot
gas. Here we investigate the effects that deviations from equilibrium
have on $M$ and $\beta$ estimates, by using simple analytical
calculations and hydrodynamical simulations representative of gas rich
galaxies.  We show that the deviations of the X-ray determined
$\beta^{est}$ and $M^{est}$ from the true values are linked by a
remarkably simple relation; in particular, $M$ is underestimated if
$\beta$ is overestimated. Also, more radially anisotropic orbital
distributions than true are deduced in presence of gas infall
velocities of the order of the local stellar velocity dispersion (as
are likely in the central regions of galactic cooling flows).  The
results of this analysis are applied to the most thoroughly
investigated bright elliptical, NGC4472. First we show that 
$\beta^{est}$ recently derived from X-rays corresponds to a galaxy
that is unstable by radial orbit instability. Then, assuming as true
$\beta$ and $M$ the optically derived values, we show that the
differences $\beta^{est}-\beta$ and $M^{est}-M$ agree with the
predictions found here in the case of lack of hydrostatic equilibrium,
which points to the latter as a possible explanation for the
discrepancies. 
This analysis casts doubts on the possibility to use
X-ray information to determine accurately the dynamical properties of
bright X-ray emitting ellipticals, at least within their $\Reff$.

\end{abstract}

\begin{keywords}
galaxies: elliptical and lenticular, CD -- galaxies: kinematics and
dynamics -- galaxies: structure -- galaxies: cooling flows -- X-rays:
galaxies -- X-rays: ISM
\end{keywords}

\section{Introduction}

Since its discovery in the late 1970's (Fabbiano 1989) the presence of
an extended and X-ray emitting interstellar medium (ISM) in early type
galaxies raised the hope of having a new powerful and reliable tool to
determine the dynamical properties of these galaxies, in particular
their mass profile and total mass out to large radii ($\sim 10$
optical effective radii $\Reff$).  In the simplest and commonly
adopted approach, these determinations assume a spherical mass
distribution. Then, the most general expression for the momentum
equation of the gas is:
\[
{k\over \mu m_p}{1\over \rho} {d\, \rho T\over dr}+\nh=- g,
\] 
where $T$ and $\rho$ are the radial profiles of the ISM temperature
and density, $m_p$ is the proton mass, $\mu$ is the mean molecular
weight and $g=GM(r)/r^2$ is the total gravitational field. $\nh$
includes all the non hydrostatic terms and, for simplicity, we do not
consider other pressure terms such as the magnetic one (see however
Sect. 4). Thus, in hydrostatic equilibrium, $T(r)$ and $\rho (r)$ --
that in principle can be inferred from the X-ray surface brightness
profile through deprojection, using also spectral information (e.g.,
Forman et al. 1985, Trinchieri et al. 1986) -- are linked to the total
mass profile by the well known formula (Fabricant, Lecar \& Gorenstein
1980)
\[
M(r)=-r{k T\over G\mu m_p}\left( {d\,{\rm log}\rho \over d\,{\rm log} r}
+{d\,{\rm log}T \over d\,{\rm log} r} \right).
\]
Forman et al. (1985) made the first total mass determinations using
this approach, for a sample of early type galaxies with {\it Einstein}
data extending out to radii of $\sim 50$ kpc. They obtained large mass
values implying that dark matter should be present in a large quantity
($M/L_B\sim 25-100\, M_{\odot}/L_{B,\odot}$). However, uncertainties
in the $\rho$ and $T$ profiles were so large that mass values could be
even as low as those derived from optical measurements (Trinchieri,
Fabbiano \& Canizares 1986; Fabbiano 1989), that typically give
$M/L_B\sim 12 M_{\odot}/L_{B,\odot}$ within $\Reff$ (Saglia,
Bertin \& Stiavelli 1992).

Mass profiles can in fact be derived also from the observed stellar
velocity dispersion profiles $\sigma_p(R)$; these extend out to $\sim
1-2 \Reff$ in few best cases (e.g., Kronawitter et al. 2000). In this
context, a common approach is to use the Jeans equations; for a spherical
model described by a phase-space distribution function $f=f(E,L^2)$,
the relevant equation is:
\[
{1\over \rho_*} {d \rho_* \sigma _r^2\over d r}
+{2\beta  \sigma _r^2\over r}=-g
\]
(e.g., Binney \& Tremaine 1987).  Here $\rho_*$ is the stellar density
profile and $\beta(r)=1-\sigma_t^2(r)/2\sigma_r^2(r)$, where $ \sigma
_r$ is the radial stellar velocity dispersion and $ \sigma^2
_t=\sigma^2_{\varphi}+\sigma^2_{\vartheta}$ is the total tangential
stellar velocity dispersion.  $\beta(r)$ describes the radial trend of
anisotropy in the stellar velocity dispersion tensor, which is
radially anisotropic where $0<\beta<1$, tangentially anisotropic where
$\beta<0$ and isotropic where $\beta=0$.  For assumed visible and dark
density profiles and a specified radial trend of $\beta(r)$, eq. (3)
is solved to derive $\sigma ^2_r(r)$; then by projection one obtains
$\sigma_p(R)$ to be compared with the observed one.  As well known, in
absence of a physically motivated choice for $\beta(r)$, this method
suffers from a ``dark matter--$\beta$ degeneracy'': a sufficient
amount of orbital anisotropy can mimic the presence of a massive dark
matter halo (e.g., Tonry 1983).  Observationally, $\beta(r)$ can be
derived from the deviation of the line profiles from a Gaussian (e.g.,
Gerhard et al. 1998), but this process requires high quality data and
care in analysis.  In fact, the use of the hot ISM to derive dynamical
quantities was very welcomed also because it is exempt from this
degeneracy problem\footnote{Note that in other approaches (e.g.,
Bertin et al. 1994), the anisotropy profile is not a function to be
specified by independent arguments, but derives from the physical
picture on which the model is based.}.

In hydrostatic equilibrium, the two conditions (1) (with $\nh=0$) and
(3), holding for the gas and the stars lying in the same total
gravitational field $g$, allow us to determine $M(r)$ and $\beta(r)$
uniquely, when coupled with the knowledge of the four observables
$T(r)$, $\rho(r)$, $\rho_*(r)$ and $\sigma_p(R)$. For example, by
eliminating $g$ from (1) and (3) one obtains a relation linking gas
and stellar properties only (Fabricant et al. 1989):
\[
{k\over \mu m_p \rho}{d\,\rho T\over dr}={1\over \rho_*} {d\, \rho_*
\sigma _r^2\over d r}+{2\beta \sigma _r^2\over r}.
\]
Then, one can use the equation above to derive $\beta(r)$ 
using information coming from X-ray observations, as attempted recently by
Mathews \& Brighenti (2003a; hereafter MB03).

The lack of reliable ISM density and temperature profiles has been a
severe limitation to the use of X-rays as a probe of the galactic
potential and of stellar dynamical quantities. This was first overcome
partially thanks to $ROSAT$ observations.  Currently, quite accurate
$T$ and $\rho$ profiles are being derived from $Chandra$ and
$XMM-Newton$ observations (e.g., Sun et al. 2003). In addition, in a
few cases $Chandra$ has already shown signs of departure from the
hydrostatic equilibrium in the hot gas coronae (e.g., Jones et
al. 2002).  Therefore a timely question is how much reliable the X-ray
estimate of dynamical quantities can be.  

In this paper we address the problem of the accuracy of the X-ray
based total mass $M(r)$ and $\beta(r) $ determination when the
hydrostatic equilibrium hypothesis is assumed but it does not
hold. Our analysis relies upon the use of the relevant hydrodynamical
equations, while we do not discuss technical problems related with the
extraction of physical quantities from the observations, such as the
well known ill-posed problem of Abel inversion for observed data
(e.g., Tikhonov \& Goncharsky 1987).

\section{Estimates of mass and orbital anisotropy when $\nh \neq 0$}

In the most general case ($\nh \neq 0$) the physically correct 
analog of eq. (4) is
\[
{k\over \mu m_p \rho}{d\,\rho T\over dr}+\nh =
{1\over \rho_*} {d\, \rho_* \sigma _r^2\over d r}+{2\beta \sigma _r^2\over r}.
\]
Instead, by erroneously assuming hydrostatic equilibrium, one
uses an equation involving $\beta^{est}$:
\[
{k\over \mu m_p\rho}{d\rho T\over dr}={1\over \rho_*} {d\, \rho_* \sigma _r^2\over d r}+{2\beta^{est} \sigma _r^2\over r}.
\]
Therefore the difference between the estimated and the true anisotropy
profile is given by
\[
\beta^{est}-\beta=-{r\over 2\sigma_r^2}\nh.
\]
Following the same reasoning, the equation that should be used to
derive the total mass profile $M(r)$ is eq. (1), while using it with
$\nh=0$, one derives an estimated mass $M^{est}$ given by:
\[
M^{est}=M+{r^2 \over G}\nh.
\]
Note that, although $M^{est}$ and $\beta^{est}$ derive from two independent
procedures, the deviation in the mass estimate is linked to
that in the $\beta$ estimate by a remarkably simple relation:
\[
{M^{est}(r)\over M(r)}-1=-{2r\sigma_r^2(r) [\beta^{est}(r)-\beta (r)]\over
G M(r)}.
\]
Thus, for example, $M^{est}<M$ wherever $\beta^{est}>\beta$.  

In general, the modulating factor $2r\sigma_r^2/G M$ in eq. (9)
depends on the exact slope of the stellar density profile, on the
presence of dark matter and on $\beta(r)$ [via $\sigma_r(r)$]. The
radial behavior of this factor for a few representative cases is given
in Fig. 1. Within $\sim 0.2 \Reff$ the factor is very sensitive to the
shape of the density profile, as shown by the cases of the de
Vaucouleurs (1948), Jaffe (1983) and Hernquist (1990) isotropic models
(respectively the solid, long-dashed and dotted lines). In addition,
Fig. 1 also shows that this dependence is not trivially linked to the
radial behavior of the $\rho_*$ profile, since $\rho_*(r)\propto
r^{-3/4}, \,\, r^{-2} $ and $r^{-1} $ respectively for the de
Vaucouleurs (Ciotti 1991), Jaffe and Hernquist models.  The inner
shape of the factor can be derived, however, from the asymptotic
analysis discussed by Bertin et al. (2002).
 
The short-dashed line in Fig. 1 refers to an isotropic stellar de
Vaucouleurs model coupled to a dark matter halo that is described by an
Hernquist profile and is 5 times as massive as the stellar component;
the mass enclosed within $\Reff$ is the same for the two components
(which implies a core radius for the Hernquist model of $2.47
\Reff$). Compared to the corresponding pure stellar model (solid line)
there is a small increase in the modulating factor.

When $\beta> 0$ the factor is larger than in the isotropic case; for
example, when $\beta $ is radially constant\footnote{Any consideration
on the physical consistency of this assumption is left apart here;
see, e.g., Camm 1952.}, as adopted in some modelings (see, e.g.,
Sect. 3), and the density distribution is $\rho_*\propto r^{-\gamma}$,
the factor is $1/(\gamma-1-0.5\beta)$. In Fig. 1 the dot-dashed line
refers to a de Vaucouleurs profile with $\beta(r)$ radially varying
according to the Osipkov-Merritt parameterization (Osipkov 1979,
Merritt 1985).  In the central region, the effect of such a $\beta
(r)=r^2/(r^2+r_a^2)$ (where $r_a$ is the anisotropy radius) is a small
increase of the modulating factor with respect to the isotropic case
(solid line).  In the external regions an idea of the full range of
the factor values derives by comparing the cases of isotropic and
anisotropic orbital distributions for a de Vaucouleurs profile (solid
and dot-dashed lines respectively). In fact, for this profile, the
choice of $r_a\simeq \Reff$ adopted for Fig. 1 gives the maximum
allowed level of anisotropy, based on stability arguments (Ciotti \&
Lanzoni 1997; see also Sect. 3).

In summary, outside the effective radius the modulating factor is
everywhere $< 0.7$; for the ``maximally'' anisotropic case, it is
everywhere $<1$.  Within $\Reff$ it is again everywhere $ \lsim 1$,
except for the Hernquist model. Therefore for most models 
``errors'' on $M$ (i.e., $M^{est}/M-1$) are $\lsim $ than the
$\beta$ deviations (i.e., $\beta^{est}-\beta$).

In the next sections we estimate the difference
$\beta^{est}-\beta$ in eq. (7), first qualitatively (Sect. 2.1) and
then quantitatively by using numerical models of hot gas flows for 
X-ray bright galaxies (Sect. 2.2).

\subsection{Qualitative analysis}

We assume that the hot ISM filling the galactic potential well
has velocity, density and temperature 
$u$, $\rho $ and $T$.  Usually, in the case of galactic
flows, a mass source rate $\alpha \rho_*(r) $ is present in the $\nh $
term, due to the stellar mass losses (e.g., Ciotti et al. 1991).  For
an inviscid fluid, the $\nh$ term is then written as
\[
\nh ={\partial u\over \partial t} + {1\over 2}{du^2
\over dr}+{\alpha \rho_* u \over \rho}.
\]
We consider first eq. (7) in the simplest case, namely the stationary
case with $\alpha =0$; then
\[
\beta^{est}-\beta=-{r\over 4\sigma_r^2}{d u^2\over dr}.
\]
This shows that when the velocity decreases with increasing radius
{\it in magnitude}, i.e., when $du^2/dr<0$, then
$\beta^{est}(r)>\beta(r)$.  This is the case of the central regions in
cooling flows.  If, instead, the velocity increases in modulus with
increasing radius, then $\beta(r)$ is underestimated. This could be
the case of galactic winds. Note that fairly large
differences $\beta^{est}-\beta $ are expected when $u^2$ varies by
$\sim \sigma_r^2$ over a characteristic length of the order of $r$.

When $\alpha > 0$, these results are reinforced in two cases: when the
gas is inflowing ($u<0$) with a velocity decreasing with radius
($du^2/dr<0$), then even more radially anisotropic orbits are deduced;
and when the gas is outflowing ($u>0$) with a velocity increasing with
radius ($du^2/dr>0$), then even more tangentially anisotropic orbits
are deduced. In the other cases the $\alpha \rho_* /\rho$ term may
have a competing effect on the $du/dr$ term that depends on the
magnitude of the source function $\alpha$.

All the results above can be converted straightforwardly into results
concerning mass estimates by using eq. (9).

The X-ray brightest galaxies, for which better quality data are
available and dynamical studies can be attempted, contain large
quantities of hot gas and are supposed to host inflows for which
$du^2/dr<0$. In their inner regions (well within $\sim \Reff$) the gas
can reach high velocities, up to the local sound speed, while $u\sim 0$
outside $\Reff$ (e.g., Ciotti et al. 1991). Therefore, it is expected
that more radially anisotropic orbits than true are deduced from X-ray
observations of the central regions; $\beta^{est}$ derived
outside $\Reff$ should be reliable.
More quantitative estimates follow below.

\subsection{Quantitative results}

Here we study the effect of the full $\nh$ term given
in eq. (10). For this investigation, two gas dynamical models are
used, resulting from the numerical simulations of time evolving
galactic gas flows by Pellegrini \& Ciotti (1998) and representative
of hot gas rich galaxies. In principle, one should use a numerically
computed right hand side of eq. (10). However, in order to reduce the
numerical noise due to spatial and temporal derivatives over a finite
grid, the $\nh$ term is obtained from eq. (1), so that:
\[
\beta^{est}-\beta={r\over 2\sigma^2_r} \left (g+{k\over \mu m_p\rho}{d
\, \rho T \over dr}
\right).
\]
Similarly:
\[
{M^{est}\over M}=-{k\over g\mu m_p\rho}{d\, \rho T \over dr}.
\]
In fact, for the galaxy models we know $g$ and $\beta$. As
representative cases we consider 1) a bright inflowing galaxy ($L_B=7
\times 10^{10}L_{\odot}$, $\Reff=9.0$ kpc, central $\sigma=287$ km
s$^{-1}$ and a dark halo nine times as massive as the stellar
component); 2) a less massive galaxy in which the inflow is confined
within 0.1 $\Reff$ ($L_B=2.2 \times 10^{10}L_{\odot}$, $\Reff=4.4$
kpc, central $\sigma=210$ km s$^{-1}$ and a dark halo twice as massive
as the stellar component).  The underlying galaxy model is isotropic
($\beta=0$).

Fig. 2 shows in the upper panel the radial trend of the gas velocity
at the end of the simulations (corresponding to an elapsed time of 15
Gyrs).  The panel below shows whether any term in the right hand side
of eq. (10) gives a dominant contribution to $\nh$. Since the gas
flows have reached a quasi steady state after 15 Gyrs, the $\partial
u/ \partial t$ term is negligible, and only the radial trend of the
advective/source term ratio is
plotted. For the largely inflowing galaxy the advective term is 
dominating in the central region and becomes comparable to the source term 
at $\sim \Reff$, while for the partial wind model the two terms keep 
comparable. For both models the two terms produce a deviation in the same
sense, since they are both negative in the inflowing region and positive
in the outflowing region.

In the lowest two panels, Fig. 2 shows what $\beta^{est}$ and
$M^{est}$ would be derived from X-ray
observations.  Overall, the trends of $\beta^{est}$ agree with the
qualitative analysis of Sect. 2.1 and, as expected, $\beta^{est}$ is $
>0$ for the central gas inflowing region. For the galaxy hosting a
global inflow, decidedly more radially anisotropic orbits than true
would be deduced by neglecting the gas velocity; correspondingly,
$M^{est}<M$ would be derived for the central galactic region
[eq. (9) and Fig. 2].  Outside the inflow region, $\beta^{est} $ is small
and $M^{est}\sim M$,
consistently with the predictions for low gas velocities.

\begin{figure}
\hskip -0.5truecm
\includegraphics[height=.5\textheight,width=.58\textwidth]{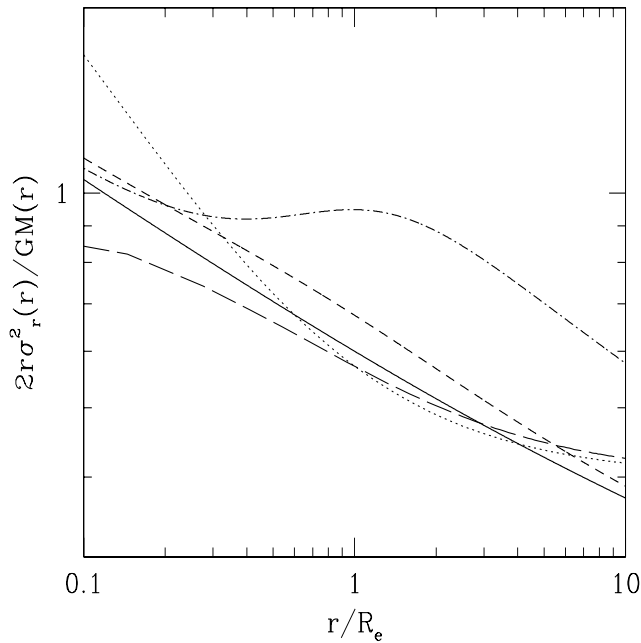}
\caption{The modulating factor of eq. (9) for the isotropic Hernquist
(dotted line), Jaffe (long-dashed line) and de Vaucouleurs (solid
line) density profiles. The dot-dashed line represents a
maximally anisotropic de Vaucouleurs model (see Sect. 2). The
short-dashed line is an isotropic stellar de Vaucouleurs model plus
a Hernquist dark matter halo five times as massive as the stellar
component; the mass enclosed within $\Reff$ is the same for the two
components (see Sect. 2).
\label{fig2}}
\end{figure}

\begin{figure}
\hskip -1.2truecm
\includegraphics[height=.55\textheight,width=.6\textwidth]{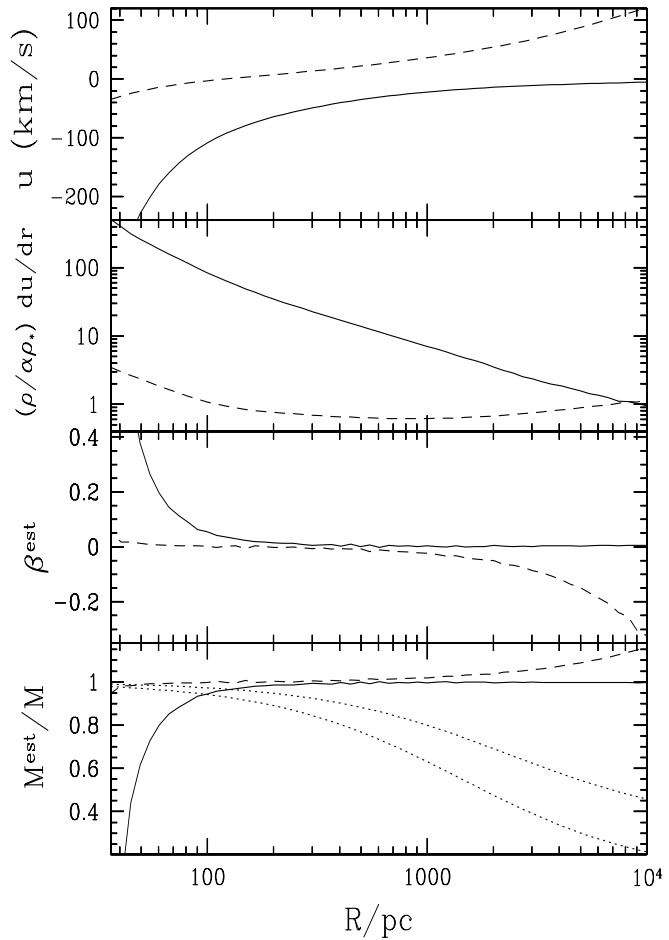}
\caption{For a global inflow (in a galaxy with $L_B=7.1 \times
10^{10}L_{\odot}$, solid line) and a partial wind (in a galaxy with
$L_B=2.2 \times 10^{10}L_{\odot}$, dashed line; more details on the
model galaxies are given in Sect. 2.2) from top to bottom the panels
show the radial trends of: the gas flow velocity, the ratio between
the advective and the source term in eq. (10), $\beta^{est}$, and
$M^{est}$ normalized to the true total mass. In the lowest panel the
dotted lines give the stellar mass profiles of the two model galaxies
(again normalized to the total mass;
see Sect. 2.2 for more details).
\label{fig1}}
\end{figure}

\section{The case of NGC4472 }

The X-ray bright Virgo galaxy NGC4472 has been chosen since the
earliest times for a detailed investigation of its dynamical
properties using X-ray data (Thomas 1986; Trinchieri et al.  1986;
Loewenstein 1992; Bertin, Pignatelli \& Saglia 1993; Bertin \&
Toniazzo 1995; Irwin \& Sarazin 1996). The study based on the most
recent and extensive data set is that of MB03 (see also Mathews \&
Brighenti 2003b). MB03 employed eq. (4) to relate $\beta(r) $ and
$T(r)$. They used $\rho(r)$ derived from Abel-inverted $Chandra$,
$ROSAT$ and $Einstein$ surface brightness profiles, $T(r)$ derived
from deprojection of $ROSAT$ and $Chandra$ data, the observed
$\sigma_p(R)$ from Bender et al. (1994) and Fried \& Illingworth
(1994) and $\rho_*(r)$ resulting from fitting the observed brightness
profile with a de Vaucouleurs law (but in the central regions the
density profile was assumed to steepen as $r^{-0.9}$).

Adopting a radially constant $\beta^{est}$ for the modeling, the value
$\beta^{est}=0.71\pm 0.15$ was derived, while the optically determined
$\beta(r)$ within $\Reff$ is everywhere $\lsim 0.34$ (Kronawitter et
al. 2000).  That is, 
$\beta^{est}$ is about twice as large as the maximum $\beta$ found
from optical data.  Conversely, assuming the optically determined
$\beta(r)$ of Kronawitter et al. (2000), $T(r)$ higher than observed
is obtained.  Commenting this result, MB03 state that ``{\it This
strong preference for radial stellar orbits must be understood in
terms of the formation history of massive elliptical
galaxies. Conversely, if the smaller, optically determined $\beta$ is
indeed correct, we are led to the important conclusion that the
temperature profile $T(r)$ of the hot ISM in NGC 4472 must differ from
that indicated by X-ray observations, or that the hot gas is not in
hydrostatic equilibrium.}''

Here we argue that, if problems with the several steps involved in the
modeling of the data can be excluded, the last possibility of lack of
hydrostatic equilibrium is strongly favoured.  In fact a simple
stability argument rules out as unphysical such a high X-ray--derived
$\beta^{est}$ value. In case of a constant anisotropy parameter, the
radial orbit instability indicator (Polyachenko \& Shukhman 1981,
Fridman \& Polyachenko 1984) turns out to be
\[
{2K_r\over K_{\vartheta}+K_{\varphi}}={1\over 1-\beta^{est}}\simeq 3.45,
\]
where $K_r$ and $K_{\vartheta}+K_{\varphi}$ are respectively the total
radial and tangential kinetic energies. A galaxy is unstable due to
radial orbit instability when this indicator is $\gsim 1.7\pm 0.25$;
therefore the value calculated in eq. (13) is well outside the
stability regime.  Note that even when inserting the maximum value of
$\beta(r)=0.34$ derived by Kronawitter et al. (2000) in eq. (13), the
indicator is $\approx 1.5$, safely in the stability regime. The
criterion above has been proved to be obeyed essentially by all galaxy
models to which it has been applied, even including massive dark halos
(e.g., Stiavelli \& Sparke 1991; Nipoti, Londrillo \& Ciotti 2002;
Bertin \& Trenti 2003; see also Saha 1991).

Additional considerations argue in favour of the hypothesis of lack of
hydrostatic equilibrium to explain the inconsistencies between X-ray
and optically-derived orbital anisotropy.  In fact, there is also a
deviation between the X-ray and the optically--derived mass for
NGC4472, and the two discrepancies (concerning $\beta$ and $M$) are
related as predicted by eq. (9), as shown in the following. Having
discarded the X-ray--derived $\beta^{est}$ as unphysical, we assume
that the optical estimate of $\beta$ is the correct one. Then, eq. (9)
predicts the X-ray--derived $M^{est}$ to underestimate the
true mass $M$. More specifically, the expected value of $M^{est}/M$ can be
derived from the results for the de Vaucouleurs models in Fig. 1.  For
a range of values of the modulating factor within $\Reff$ of $\sim
0.5-1$ and for $\beta^{est}-\beta=0.4$, then $M^{est}/M= 0.8-0.6$ is
given by eq. (9). What $M^{est}$ derives from the X-ray data?
The best fit total mass profile $M^{est}(r)$ determined from eq. (2),
with the same $T(r)$ and $\rho (r) $ used by MB03, is a one-component
de Vaucouleurs mass profile out to $r=\Reff$, with a total constant
mass-to-light ratio of $M^{est}/L_B=7$ (Mathews \& Brighenti 2003b).
$M/L_B$ estimates were derived also from optical studies based on high 
quality photometrical and kinematical
data extending out to $\sim 0.5\Reff$.
The $M/L_B$ determined from one-component constant mass-to-light ratio
axisymmetric dynamical models is 9.2, for the central galactic region,
when rescaled to the distance adopted by MB03 (van der Marel 1991).
Another analysis used spherical self-consistent two-component models
(Saglia et al. 1992). The models selected at a 3 sigma confidence
level have $M/L_B=10.1-14.6$ within $\Reff$, when rescaled to the MB03
distance\footnote{These selected models also have a velocity
dispersion profile consistent with the data for the globular cluster
system out to $\sim 2.5\Reff$.}. These optically determined $M$
values, if taken as true values, imply that
$M^{est}$ underestimates the
true mass $M$. More precisely,
$M^{est}/M= 0.76$ and $0.69-0.48$ when adopting respectively the van
der Marel's and the Saglia et al.'s results, in remarkable agreement
with the predictions above ($M^{est}/M= 0.8-0.6$).  Therefore, the
hypotheses of i) correct optical estimates of $\beta$ and $M$ and ii)
departure from hydrostatic equilibrium of the hot ISM, give a
consistent explanation for the discrepancies of the X-ray--derived
$\beta^{est} $ and $M^{est}$ values. In particular, these hypotheses
account for the fact that $\beta^{est}$ is an overestimate and
$M^{est}$ is an underestimate and also for the {\it size} of the
discrepancies with the true $\beta$ and $M$.

\section{Discussion and Conclusions}


We have investigated the effect that deviations from hydrostatic
equilibrium have on the X-ray estimates of the total mass $M$ and the
orbital anisotropy parameter $\beta$ via the ``standard'' eqs. (2) and
(4). We have found that:

$\bullet $ in the most general case, the deviations of $\beta^{est}$
and $M^{est}$ from the true values are linked by a remarkably simple
relation. In particular, the mass is underestimated if $\beta$ is
overestimated. Also, for the most common density and orbital
anisotropy distributions, with or without a massive dark halo, the
$\beta^{est}$ deviations are larger than the ``errors'' on $M$ at all
radii $\gsim 0.2\Reff$.

$\bullet $ in the stationary case without mass/momentum sources/sinks,
more radially anisotropic orbits than true are deduced for the central
regions of gas inflowing galaxies.
Instead, $\beta(r)$ is underestimated for outflowing regions where the
velocity increases with radius, as in galactic winds.
When mass sources are present, the results above are
reinforced in the inflow and in the wind cases. In the
other cases the mass source term may have a competing effect.

$\bullet $ the full time dependent problem is investigated with the
aid of numerical simulations reproducing gas rich galaxies.  The
results are consistent with the simple qualitative predictions.
$\beta^{est}$ decidedly overestimates the true value (and $M^{est}$
underestimates it) if derived from the central region where gas is
inflowing with a large velocity, while outside it $\beta^{est} \sim
\beta $ and $M^{est}\sim M$.

$\bullet$ $\beta^{est}$ and $M^{est}$ derived for the most thoroughly
investigated bright elliptical galaxy (NGC4472), for which the best
X-ray data are available, are considered as a representative
application of our results.  We show that the high $\beta^{est}$
recently derived is unphysical. We also note a discrepancy in the
X-ray and optically derived masses within $\Reff$. Assuming as true
$\beta$ and $M$ the optically derived values, the $\beta$-overestimate
implies (when applying our results for the non-hydrostatic
equilibrium) an $M$-underestimate in remarkable agreement with
$M^{est}-M$.  Of course, one cannot exclude that a different modeling
of the X-ray and optical data would be able to produce a $\beta(r)$
profile consistent with the observed one, in the hydrostatic
equilibrium hypothesis.  However, this would reveal a degeneracy in
the procedure and undermine the possibility to uniquely derive
dynamical information from the X-ray data.

$\bullet $ note that eqs. (7--9) describe the deviations of
$\beta^{est}$ and $M^{est}$ from the true values even when sources of
pressure additional to the thermal one are effective (such as magnetic
and radiation pressure). In fact, if these are accounted for
explicitely in eqs.(2) and (4), then eqs. (7--9) remain unaltered. If
these are not accounted for, again eq. (9) remains unaltered and 
eqs.(7--8) hold provided that the additional pressure terms are
considered as included in the $\nh$ term.
The deviations of $\beta^{est}$ and $M^{est}$ produced by these
kinds of pressure may be confined to the central galactic regions,
 where magnetic fields and AGN-related phenomena more likely
are concentrated.

Additional explanations can be figured out for the deviations of
X-ray--derived $\beta^{est}$ and $M^{est}$ from the true values.
These include deviations from spherical symmetry,
projection/deprojection effects not adequately taken into account,
presence of a multiphase ISM at each radius,
amplification of uncertainties in the process of extracting $T(r)$,
$\rho(r)$, $\rho_*(r)$ and $\sigma_r(r)$ from the data.  However, it
is yet to be investigated the possibility for these explanations to
account at the same time for deviations in $\beta$ and $M$, and by
precisely the observed amount, as demonstrated here for the hypothesis
of lack of hydrostatic equilibrium in the case of NGC4472.

Our analysis casts doubts on the possibility to use X-ray information and
the hydrostatic equilibrium assumption as robust tools to determine
accurately the dynamical properties of X-ray emitting galaxies, at
least within $\Reff$. Another
 warning to take with care X-ray mass estimates
had already been made when it appeared that mass models
constrained by accurate and extended stellar dynamical data, and
including different amounts of dark matter, all lead to similar
temperature and X-ray brightness profiles (Bertin et al. 1993), within
the uncertainties in the X-ray measurements. In this sense, the
situation might be improving with the new $Chandra$ and $XMM$ data.
 
In conclusion, instead of obtaining robust dynamical estimates from
hot gas-based studies, it should perhaps be expected that stellar
dynamical studies (possibly coupled with gravitational lensing
studies) help understand the dynamical status of the hot gas.

\section*{Acknowledgments}
We would like to thank G. Bertin for useful discussions and the referee,
M. Loewenstein, for comments that improved the paper.

\end{document}